\documentclass[aps,prb,twocolumn,superscriptaddress,floatfix]{revtex4-2}
\usepackage{graphicx,graphics}
\usepackage{natbib}
\usepackage{lipsum}
\usepackage{dcolumn}
\usepackage{amsmath,amssymb,amsfonts}
\usepackage{latexsym,verbatim}
\usepackage{bm}
\usepackage{color}
\usepackage{ulem}
\usepackage{siunitx}
\usepackage[breaklinks=true,colorlinks,citecolor=blue,linkcolor=blue,urlcolor=blue]{hyperref}

\begin{document}
	
\title{Dynamical polarizability of graphene with spatial dispersion}
\author{Tao Zhu}
\email{phyzht@nus.edu.sg}
\affiliation{Department of Physics, National University of Singapore, Singapore 117551, Republic of Singapore}
\author{Mauro Antezza}
\affiliation{Laboratoire Charles Coulomb (L2C), UMR 5221 CNRS-Universit\'{e} de Montpellier, F-34095 Montpellier, France}
\affiliation{Institut Universitaire de France, 1 rue Descartes, F-75231 Paris, France}
\author{Jian-Sheng Wang}
\affiliation{Department of Physics, National University of Singapore, Singapore 117551, Republic of Singapore}

\begin{abstract}
We perform a detailed analysis of electronic polarizability of graphene with different theoretical approaches. From Kubo's linear response formalism, we give a general expression of frequency and wave-vector dependent polarizability within the random phase approximation. Four theoretical approaches have been applied to the single-layer graphene and their differences are on the band-overlap of wavefunctions. By comparing with the \textit{ab initio} calculation, we discuss the validity of methods used in literature. Our results show that the tight-binding method is as good as the time-demanding \textit{ab initio} approach in calculating the polarizability of graphene. Moreover, due to the special Dirac-cone band structure of graphene, the Dirac model reproduces results of the tight-binding method for energy smaller than \SI{3}{\electronvolt}. For doped graphene, the intra-band transitions dominate at low energies and can be described by the Lindhard formula for two-dimensional electron gases. At zero temperature and long-wavelength limit, with the relaxation time approximation, all theoretical methods reduce to a long-wave analytical formula and the intra-band contributions agree to the Drude polarizability of graphene. Effects of electrical doping and temperature are also discussed. This work may provide a solid reference for researches and applications of the screening effect of graphene.
\end{abstract}
\maketitle

\section{Introduction}

As the first isolated single-layer material, graphene has attracted intense interest for its unique electronic structures \cite{graphene,graphene2,graphene3}. Due to the unusual gapless linear band structure with a vanishing density of state at the Fermi energy, screening properties in graphene differ significantly from conventional two-dimensional (2D) materials and become one of the most important quantities for prospective applications in many fundamental physics \cite{graphene4}. The dynamical electric screening can be charactered by a frequency and wave-vector dependent polarizability $\Pi({\bf q},\omega)$ which describes the response of induced charge density to a time-dependent effective potential. The quantities such as electrical conductivity, dielectric function, electrical susceptibility, reflection coefficient, have simple relations with the electronic polarizability that usually involve the pure Coulomb interaction $v$. For example, the electrical dielectric function has the form $\varepsilon = 1-v\Pi$ with its zeros correspond to the poles of the response function that describe the plasmon modes. In the momentum space, the 2D electrical conductivity has the relation $\sigma = i\omega\Pi/q^2$ that can be obtained from the linear response electrodynamics. These quantities have been proved to play a pivotal role in developing a variety range of physical applications especially for the near-field energy transport \cite{energy1,energy2,energy3,energy4,energy5,energy6,energy7,energy8} and the Casimir effects \cite{casimir1,casimir2,casimir3,casimir4,casimir5,casimir6}.

In general, the polarizability of graphene is nonlocal that depends on both the frequency and the wave-vector. Other factors like temperature, electrical doping (chemical potential), and sometimes the mass-gap can also play an important role. Many theoretical approaches have been developed to study the screening effect of graphene, for both polarizability \cite{wunsch,huang} and conductivity \cite{gusynin,gusynin2,falkovsky,falkovsky2,klimchitskaya,klimchitskaya2}. In particular, Wunsch \textit{et al} \cite{wunsch} gives an analytical formula of dynamical polarizability with the Matsubara technique at zero temperature, which agrees with the work of Hwang and Das Sarma \cite{huang}. In the long-wavelength limit $q\to0$, Falkovsky and Varlamov \cite{falkovsky,falkovsky2} obtained simple analytical expressions of conductivity of graphene via Green's functions approach. Besides, Bordag \textit{et al} \cite{bordag1,bordag2} performed the quantum electrodynamics in (2+1)-dimensional space-time to get the complete components of polarization tensor of graphene. Then the electrical conductivity of graphene at arbitrary temperature is developed taking into account mass-gap parameter by Klimchitskaya \textit{et al} \cite{klimchitskaya,klimchitskaya2}. However, most of the theoretical treatments are based on the Dirac model that the energy dispersion is assumed to be linear around the K points of the Brillouin zone. This assumption is only applicable to the nearest $\pi$-band of graphene with energy less than a few electron volts. Moreover, with finite electrical doping, the Drude model has been widely used to calculate the intra-band contributions to the heat transfer and the Casimir force \cite{energy1,casimir1}. A detailed discussion on the validity of these theoretical treatments is still lacking. 

In this article, we present a derivation, comparison, and discussion of different methods to calculate the frequency- and wave-vector-dependent electric polarizability of graphene. The effects from temperature and chemical potential are also discussed. We focus on the density-density correlation function of graphene at the quasi-static limit such that only the longitudinal component of polarizability is considered in this work. We also do not consider the dependence on the mass gap parameter throughout our discussion. We study four theoretical approaches, namely, density functional theory (DFT) based \textit{ab initio} method, nearest-neighbor tight-binding method, the Dirac model, as well as the Lindhard formula for two-dimensional electron gas (2DEG) \cite{lindhard,mahan}. From the linear response Kubo formula \cite{kubo}, we obtain a general expression of the electrical polarizability and then employ it to graphene with different approximations. The only difference between different methods is given by a so-called prefactor that describes the band-overlap of wavefunctions. Our results show that the tight-binding method can produce comparable results to the \textit{ab initio} calculation. When the energy is smaller than 3 eV, the Dirac model reproduces the results of the tight-binding method. With finite electrical doping, intra-band transitions become dominant at low energy and can be well described by the Lindhard formula. 

\section{Polarizability}

To get explicit response function from a microscopic electronic structure, we consider a system under small time-dependent external perturbation that the Hamiltonian has the form $H = H_0+H_1(t)$, where $H_0$ is the unperturbed term and $H_1$ represents time-dependent interactions between the charge density $\rho({\bf r})$ and an external electric potential $\phi({\bf r},t)$. Assuming the perturbation is switched on adiabatically, we can write the interaction term as
\begin{equation}
	\label{H_1}
	H_1(t)=\int d^3{\bf r}\ \hat{\rho}({\bf r})\phi({\bf r},t)e^{\eta t},
\end{equation}
where $\hat{\rho}$ is the charge density operator. The damping factor $\eta$ is an infinitesimal positive quantity which ensures that the perturbation vanishes at $t \to -\infty$ and accounts for the carrier relaxation that is provided by the long-range scatterers \cite{falkovsky2}. In the interaction picture, the charge density operator $\hat{\rho}({\bf r},t) = e^{iH_0t/\hbar}\hat{\rho}({\bf r})e^{-iH_0t/\hbar}$ and the expectation value of the induced charge density $\delta \rho$ is a linear response of the external perturbation:
\begin{equation}
	\label{rho}
	\langle \delta \hat{\rho}({\bf r},t)\rangle=\int dt^\prime\int d^3{\bf r}^\prime \chi({\bf r},{\bf r}^\prime,t-t^\prime)\phi({\bf r}^\prime,t^\prime),
\end{equation}
where $\chi$ is the electric susceptibility that is given by the Kubo formula \cite{kubo}
\begin{equation}
\chi({\bf r},{\bf r}^\prime,t-t^\prime)= -\dfrac{i}{\hbar}\Theta(t-t^\prime)\langle\left[\hat{\rho}({\bf r},t),\hat{\rho}({\bf r}^\prime,t^\prime)\right]\rangle.
\label{kubo}
\end{equation}
where $\Theta$ is the Heaviside step function that equals 1 for $t>t^\prime$ and 0 otherwise.

Here we are more interested in the electric polarizability $\Pi$ which represents the linear response of the induced charge density to the total (external + induced) potential of the system. The relation between $\Pi$ and $\chi$ is given by the Dyson equation $\chi = \Pi + \Pi v\chi$ where $v$ is the bare Coulomb interactions of 2D system \cite{dyson,Onida}. For non-interacting electron systems, we can consider only the bare bubble diagram of the Dyson equation so that the electric polarizability in the random phase approximation (RPA) \cite{rpa} writes
\begin{equation}
	\label{pi}
\Pi({\bf r},{\bf r}^\prime,\omega)=2e^2\sum_{i, j}(f_j-f_i)\dfrac{\psi_j({\bf r}) \psi^*_j({\bf r}^\prime) \psi^*_i({\bf r}) \psi_i({\bf r}^\prime)}{\epsilon_j-\epsilon_i-\hbar\omega-i\eta},
\end{equation}
where we transformed Eq.~(\ref{kubo}) into frequency domain and calculate the expectation value of density operator commutator with independent particle wave-functions $\psi$ and corresponding energies $\epsilon$ of unperturbed Hamiltonian $H_0$. The index $i,j$ denote different states, the factor 2 accounts for the spin degeneracy, and $e$ is the electron charge. The function $f = (1+e^{\beta(\epsilon-\mu)})^{-1}$ is the Fermi distribution function with $\beta = 1/k_BT$ and $\mu$ the chemical potential.

For systems with a periodic crystal, it is more convenient to calculate the polarizability in momentum space. According to the Bloch theorem, the sum over states in Eq.~(\ref{pi}) can be replaced by a sum over bands and k-points ($i\to n{\bf k},j\to n^\prime{\bf k}^\prime$) where {\bf k} lies within the first Brillouin zone. Due to the lattice translation symmetry, we have $\Pi ({\bf r}+{\bf R},{\bf r}^\prime+{\bf R}) = \Pi({\bf r},{\bf r}^\prime)$, where ${\bf R}$ is any real space lattice vector. It can be shown from Fourier transform of Eq.~(\ref{pi}) that $\Pi({\bf q},{\bf q}^\prime,\omega)$ is only nonzero if ${\bf q}$ and ${\bf q}^\prime$ differ by a reciprocal lattice vector ${\bf G}$ and the quasi-momentum conservation holds, i.e. ${\bf k}^\prime = {\bf k} + {\bf q}$ modulo \textbf{G}. Writing wave-function $\psi_{n,{\bf k}}({\bf r}) = u_{n,{\bf k}}({\bf r})e^{i{\bf k}\cdot{\bf r}}$ with the Bloch function $u_{n,{\bf k}}({\bf r})$, we have 

\begin{widetext}
\begin{equation}
	\label{pi2}
	\Pi_{{\bf G},{\bf G^\prime}}({\bf q},\omega)=\dfrac{2e^2}{N\Omega}\sum_{n, n^\prime,{\bf k}}\langle u_{n{\bf k}}|e^{-i{\bf G} \cdot {\bf r}}|u_{n^\prime,{\bf k}+{\bf q}}\rangle \langle u_{n^\prime,{\bf k}+{\bf q}}|e^{i{\bf G}^\prime \cdot {\bf r}^\prime}|u_{n{\bf k}}\rangle \dfrac{f_{n^\prime,{\bf k}+{\bf q}}-f_{n,{\bf k}}}{\epsilon_{n'{\bf k}+{\bf q}}-\epsilon_{n{\bf k}}-\hbar \omega-i\eta},
\end{equation}
\end{widetext}
where $N$ is the number of k-points in the first Brillouin zone and ${\bf r}$ is the electron position operator. For 2D materials, $\Omega$ is the area of the primitive cell and ${\bf q}$ is the in-plane Bloch wave-vector.

\section{Models for graphene}
The general formula of polarizability $\Pi$ in Eq.~(\ref{pi}) and Eq.~(\ref{pi2}) are summing over transitions from different states and the key quantity is the unperturbed stationary eigenfunctions and corresponding eigenvalues. In this section, we apply this general formula to graphene with specific band structures obtained from different theoretical frameworks and approximations.

\subsection{\textit{ab initio}}

We can calculate the electronic band structure of graphene from the \textit{ab initio} methods that solve the many-body Schr\"odinger equation in the self-consistent-field approach. In principle, many first principles methods, such as the GW method \cite{gw} and the time-dependent density functional theory \cite{Onida}, can be used to obtain the excitation properties. In this work, we calculate the polarizability of graphene using DFT in the Kohn-Sham scheme \cite{ks} that the effective Hamiltonian writes
\begin{equation}
	H = \dfrac{-\hbar^2}{2m}\nabla^2+(V_{ext}+V_H+V_{xc}),
\end{equation} 
where $V_{ext}$ is the external potential, $V_H$ the Hartree potential, and $V_{xc}$ the exchange-correlation potential. The Kohn-Sham equation can be solved self-consistently to get the ground state wavefunctions which have the form of Slater determinant \cite{slater} of single-particle orbitals. In RPA, the independent particle polarizability yields the Adler-Wiser formula \cite{adler,wiser}
\begin{widetext}
\begin{equation}
\label{pi_ab}
\Pi_{{\bf G},{\bf G}^\prime}({\bf q},\omega)=\dfrac{2e^2}{N\Omega}\sum_{n,n',{\bf k}}\langle\phi_{n{\bf k}}|e^{-i({\bf q}+{\bf G})\cdot {\bf r}}|\phi_{n'{\bf k}+{\bf q}}\rangle\langle\phi_{n'{\bf k}+{\bf q}}|e^{i({\bf q}+{\bf G}^\prime)\cdot  {\bf r}^\prime}|\phi_{n{\bf k}}\rangle\dfrac{f_{n'{\bf k}+{\bf q}}-f_{n{\bf k}}}{\epsilon_{n'{\bf k}+{\bf q}}-\epsilon_{n{\bf k}}-\hbar \omega-i\eta},
\end{equation}
\end{widetext}
where $\phi_{n{\bf k}}$ and $\epsilon_{n{\bf k}}$ are Kohn-Sham wavefunctions and eigenvalues, respectively. This formula reduces to Eq.~(\ref{pi2}) if writing the Kohn-Sham wavefunctions in the form of Bloch functions, i.e. $\phi_{n,{\bf k}}({\bf r}) = u_{n,{\bf k}}({\bf r})e^{i{\bf k}\cdot{\bf r}}$. Because the ground state electronic structure corresponds to zero temperature, the Fermi function becomes a step function that $f$ equals 1 for occupied states and 0 for unoccupied states.

It is worth noting that Eq.~(\ref{pi_ab}) is not limited to the Kohn-Sham-RPA scheme and further improvements are possible. For example, one can substitute the Kohn-Sham eigenvalues $\epsilon_{n{\bf k}}$ by the quasiparticle particle energies $E_{n,{\bf k}}$ obtained from GW method (GW-RPA) as the former has intrinsic problems of underestimating the energy states. Moreover, the electron-hole interactions can be taken into account by solving the Bethe-Salpeter equation \cite{bse} on top of the GW-RPA calculations \cite{Onida}.

\subsection{Tight-binding}
 \begin{figure}
	\includegraphics[width=8.6 cm]{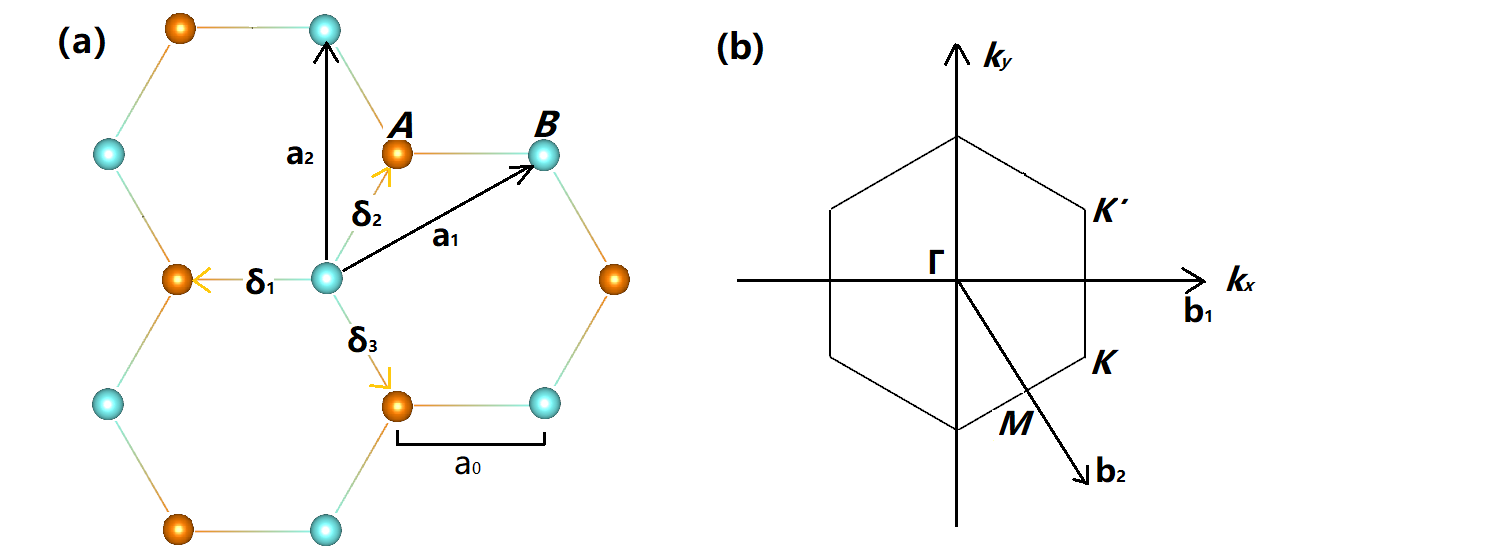}
	\caption{Honeycomb lattice of graphene and its Brillouin zone. (a) Lattice structure of graphene, $a_0 =$ \SI{1.42}{\angstrom} the distance between two nearest neighbor carbon atoms; ${\bf a}_1$ and ${\bf a}_2$ are lattice vectors of the unit cell which consists of two different sites A and B; $\delta_i,i = 1,2,3$ are the nearest-neighbor vectors. (b) The Brillouin zone of graphene. ${\bf b}_1$ and ${\bf b}_2$ are reciprocal vectors. The Dirac cones locate at the K and K$^\prime$ points.}
	\label{fig1}
\end{figure} 
Figure \ref{fig1} shows the hexagonal lattice structure of graphene and its Brillouin zone. The unit cell of graphene consists of two different sites $A$ and $B$. Considering only nearest-neighbor hopping, the tight-binding Hamiltonian for graphene has the form \cite{graphene4}
\begin{equation}
	H=-\gamma\sum_{\left<i, j\right>}(a^\dagger_ib_j+b^\dagger_ja_i),
\end{equation}  
where $a_i$ ($a^\dagger_i)$ annihilates (creates) an electron on site ${\bf r}_i$ on sublattice A (similarly for $b_j$, $b^\dagger_j$ on sublattice B). $\gamma\approx \SI{2.8}{\electronvolt}$ is the nearest-neighbor hopping parameter between different sublattices. The sum index $\left<i,j\right>$ means the nearest neighbors are summed once for each pair. 

Due to the honeycomb lattice of carbon atoms, the matrix representation of Hamiltonian of graphene is diagonal in the momentum space
\begin{equation}
	H(\bf k)={\left[ \begin{array}{ccc}
			0 & -\gamma f({\bf k})\\
			-\gamma f^*({\bf k}) & 0
		\end{array} 
		\right ]},
	\label{hamiltonian_tb}
\end{equation}
 where $f({\bf k})=\sum_ie^{i{\bf k}\cdot \delta_i}$ with $\delta_1 = a_0(-1,0)^T$, $\delta_2 = a_0(1/2,\sqrt{3}/2)^T$, and $\delta_3 = a_0(1/2,-\sqrt{3}/2)^T$ as depicted in Fig.~\ref{fig1}. The eigenvalues of the tight-binding Hamiltonian can be written as $\epsilon_{n,{\bf k}} = n\gamma |f({\bf k})|$ where $n = \pm1$ denotes two modes of the conduction ($+1$) and valence bands ($-1$). Introducing the phase factor $e^{-i\varphi({\bf k})} = f({\bf k})/|f({\bf k})|$, the cell normalized Bloch eigenstates of graphene become 

\begin{eqnarray}
	\label{eivenstates_tb}
	\nonumber S_{n,{\bf k}}&=&\frac{1}{\sqrt{2}}\left[\begin{array}{ccc}
		&1\\
		&ne^{i\varphi(\bf k)}
	\end{array}\right].
\end{eqnarray}
Because the charge distribution is homogeneous for the two sub-lattice of graphene, one can neglect the reciprocal lattice vector ${\bf G}$ (i.e., the local field effect \cite{lfe,lfe1}) and finally write the tight-binding polarizability of graphene as \cite{energy3}
\begin{widetext}
	\begin{equation}
		\label{pi_tb}
		\Pi({\bf q},\omega)=\dfrac{2e^2}{N\Omega}\sum_{n, n^\prime = \pm1,{\bf k}}S_{n,{\bf k}}^\dagger S_{n^\prime,{\bf k}+{\bf q}}S_{n^\prime,{\bf k}+{\bf q}}^\dagger S_{n,{\bf k}}\dfrac{f_{n^\prime,{\bf k}+{\bf q}}-f_{n,{\bf k}}}{\epsilon_{n'{\bf k}+{\bf q}}-\epsilon_{n{\bf k}}-\hbar \omega-i\eta}.
	\end{equation}
\end{widetext}

\subsection{Dirac model}
We can numerically calculate the tight-binding polarizability of Eq.~(\ref{pi_tb}) with proper sampling of ${\bf k}$ and ${\bf q}$ in the first Brillouin zone. However, due to the special electronic band structure, the energy dispersion of graphene is linear around the vicinity of K and K$^\prime$ points of the Brillouin zone. When ${\bf q}$ is small, we can expand the $f({\bf k})$ around the Dirac point and the matrix element of Eq.~(\ref{hamiltonian_tb}) $f({\bf k}) \approx -3a_0(k_x - i k_y)/2$ in the Dirac model. The Dirac Hamiltonian of graphene has the form $H = \hbar v_F {\bf k}\cdot \bm{\sigma}$ where $\sigma_i$ are the Pauli matrices. The Fermi velocity $v_F = 3a_0\gamma/2\hbar$ has the value $\sim10^6$ m/s that does not depend on the energy or momentum. The eigenvalues of the Dirac Hamiltonian can be expressed as $\epsilon_{n{\bf k}} =n\hbar v_F |{\bf k}|$. Substituting the eigenvectors in Eq.~(\ref{eivenstates_tb}) into Eq.~(\ref{pi_tb}) and taking into account the valley degeneracy, we can further write the Dirac polarizability of graphene as
\begin{widetext}
	\begin{equation}
		\label{pi_dirac}
		\Pi({\bf q},\omega)=4e^2\sum_{n, n^\prime=\pm1}\int\!\dfrac{d^2{\bf k}}{(2\pi)^2}\frac{1}{2}(1+nn^\prime \cos\theta) \dfrac{f_{n^\prime,{\bf k}+{\bf q}}-f_{n,{\bf k}}}{\epsilon_{n'{\bf k}+{\bf q}}-\epsilon_{n{\bf k}}-\hbar \omega-i\eta},
	\end{equation}
\end{widetext}
where we have replaced the sum of ${\bf k}$ points in the first Brillouin zone by an integral of ${\bf k}$ in the momentum space because the Dirac Brillouin zone $2\pi/a_0\to\infty$. The angle $\theta=\varphi({\bf k}+{\bf q})-\varphi({\bf k})$, and the factor 4 accounts for both spin and valley degeneracies. In the Dirac model, the phase factor has the form $e^{i\varphi({\bf k})} = (k_x+ik_y)/|{\bf k}|$ and thus $\theta$ is just the angle between ${\bf k}$ and ${\bf k}+{\bf q}$ in the momentum space. 

It is worthwhile to emphasize that Eq.~(\ref{pi_dirac}) depends on wave-vector, frequency, chemical potential, as well as temperature. This formula can be evaluated numerically by integrating the Brillouin zone with
\begin{equation}
	\cos\theta=\dfrac{{\bf k}\cdot({\bf k}+{\bf q})}{|{\bf k}||{\bf k}+{\bf q}|}
\end{equation}
for all non-vanishing wave-vector of ${\bf k}$ and ${\bf k}+{\bf q}$.

Besides the general expression of Dirac polarizability of Eq.~(\ref{pi_dirac}), many analytical formulas based on the Dirac model have been derived under certain conditions and approximations. Firstly, at zero temperature, the Fermi function becomes a simple step function that the conduction band is empty ($f_{+1}=0$) and the valence band is fully occupied ($f_{-1}=1$). For intrinsic undoped graphene, because of the gapless Dirac band structure, only inter-band transitions are possible. However, real graphene samples are usually doped that both intra- and inter-band transitions can occur. The doping effect is characterized by a nonzero chemical potential $\mu$ in the Fermi function. At zero temperature, Wunsch \textit{et al} \cite{wunsch} reported a relatively simple analytical formula of the polarizability of graphene with arbitrary wave-vectors. Similar work has been done by Hwang and Das Sarma \cite{huang} and their formula produce exactly the same numerical results as that of Wunsch's. Secondly, in the optical region, the magnitude of the wave-vector $q\to0$ and one can omit the spatial dispersion. This was done by Falkovsky \cite{falkovsky2} in the study of conductivity of graphene. Moreover, within the framework of quantum electrodynamics, Klimchitskaya \textit{et al} \cite{klimchitskaya,klimchitskaya2} report an analytical formula of conductivity of graphene based on the Dirac model at finite temperature and mass-gap.

If the scattering process induced carrier relaxations are taken into account with the relaxation time approximation ($\hbar\omega\to\hbar\omega+i\eta$) \cite{mermin,dassarma}, for the local ($q\to0$) polarizability of graphene at zero temperature $(T\to0)$ and zero mass-gap with $\omega > v_F q$ fixed, the consensus was reached for the Dirac polarizability obtained form the Kubo formalism and from the quantum electrodynamics. All reported analytical formulas reduce to what we call the long-wave polarizability of graphene \cite{wunsch,huang,falkovsky,falkovsky2,klimchitskaya,klimchitskaya2}
\begin{widetext}
\begin{equation}
	\label{pi_longwave}
	\Pi_{(T,q)\to0}(\omega)=\dfrac{ q^2e^2}{2\pi(\hbar\omega+i\eta)}\left[\dfrac{2\mu}{\hbar\omega+i\eta}+\frac{1}{2}\ln\dfrac{|2\mu-\hbar\omega-i\eta|}{|2\mu+\hbar\omega+i\eta|}-i\frac{\pi}{2}\Theta(\hbar\omega-2\mu)\right].
\end{equation}	
\end{widetext}

The first term in the square brackets of Eq.~(\ref{pi_longwave}) gives intra-band transitions that is proportional to the chemical potential. For graphene, the chemical potential can be written as $\mu = \hbar v_Fk_F$ with the Fermi momentum $k_F = \sqrt{\pi|n|}$ and $|n|$ the 2D carrier density \cite{huang}. Given the relation $\sigma=i\omega\Pi/q^2$ for 2D systems, the first term of Eq.~(\ref{pi_longwave}) yields $\sigma(\omega)=iD/\pi(\omega+i\Gamma)$ which is exactly the Drude conductivity of graphene \cite{drude}. The scattering rate $\Gamma = \eta/\hbar$ and the Drude weight $D = (v_Fe^2/\hbar)\sqrt{\pi|n|}$. 

The second and third terms represent inter-band transtions which become negligible when $\hbar\omega << 2\mu$. Nevertheless, for $\hbar\omega>2\mu$,  the inter-band transitions contribute and the corresponding conductivity of graphene has the form of $\sigma_0 = e^2/4\hbar$ which is the so-called universal conductivity of graphene \cite{klimchitskaya,Lewkowicz}.

\subsection{Lindhard}
We have shown that at finite electrical doping, the transport property of graphene is governed by the intra-band transitions when $\hbar\omega<<2\mu$. With large carrier concentrations, one can infer that the intra-band contributions to the polarizability of graphene correspond to that of 2DEG. In this case, the wavefunction has the form of plane-waves so that the cell normalized Bloch function $u_{n,{\bf k}}$ become a constant and the overlapping term of wavefunctions can be omitted. Then we have the Lindhard polarizability of 2DEG as \cite{mahan}

\begin{equation}
	\label{pi_lindhard}
	\Pi({\bf q},\omega)=4e^2\sum_{n, n^\prime}\int\!\dfrac{d^2{\bf k}}{(2\pi)^2} \dfrac{f_{n^\prime,{\bf k}+{\bf q}}-f_{n,{\bf k}}}{\epsilon_{n'{\bf k}+{\bf q}}-\epsilon_{n{\bf k}}-\hbar \omega-i\eta}.
\end{equation}

With the Dirac dispersion $\epsilon_{n{\bf k}}=\hbar v_F|{\bf k}|$, one can numerically solve Eq.~(\ref{pi_lindhard}) to get the intra-band polarizability of graphene. Interestingly, Eq.~(\ref{pi_lindhard}) corresponding to the Dirac polarizability of Eq.~(\ref{pi_dirac}) with $\theta=0$ fixed that implies ${\bf k}+{\bf q}$ has the same direction as ${\bf k}$ for intra-band transitions.

For small wave-vector $q$, we can expand the numerator of Eq.~(\ref{pi_lindhard}) as ${\bf q}\cdot\nabla_{\bf k} f_{\bf k}$ and the difference of energies in the denominator as $\hbar {\bf q}\cdot{\bf v_k} $ with ${\bf v_k}=v_F{\bf k}/|{\bf k}|$ is the carrier velocity. Then we get the analytical formula of polarizability of graphene given by Svintsov and Ryzhii \cite{svintsov} 
\begin{equation}
	\label{rpa4}
	\Pi({\bf q},\omega)=\dfrac{2e^2}{\pi\hbar^2 v_F^2}\left(\dfrac{\ln(1+e^{\beta\mu})}{\beta}\right)\left(\dfrac{s}{\sqrt{s^2-1}}-1\right),
\end{equation}
where $\beta = 1/k_BT$ and $s = (\hbar\omega+i\eta)/\hbar v_F q$. 

In Eq.~(\ref{rpa4}), when $T\to0$, the first bracket term reduces to $\mu$ and when $q\to0$, the second bracket term reduces to $1/2s^2$. In this case, we may write
\begin{equation}
	\label{pi_intra}
	\Pi_{(T,q)\to0}(\omega)=\dfrac{\mu q^2e^2}{\pi (\hbar\omega+i\eta)^2},
\end{equation}
which is exactly the Drude term of polarizability as shown in Eq.~(\ref{pi_longwave}).

\subsection{Summary}
We have derived expressions of the polarizability of graphene from different methods. In the linear response scheme, the general formula of polarizability is given by Eq.~(\ref{pi2}) that involves the eigenvalues and eigenfunctions of unperturbed Hamiltonian. Due to the special electronic properties of graphene, different theoretical frameworks and approximations can be applied at certain conditions. Up to now, we have shown four theoretical approaches, i.e., \textit{ab initio}, tight-binding, Dirac, and the Lindhard formula for intra-band transitions. For all of these methods, we may write a unified formula of the polarizability of graphene as:

\begin{equation}
	\label{pi_summary}
	\Pi({\bf q},\omega)=2e^2\sum_{n, n^\prime}\int\!\dfrac{d^2{\bf k}}{(2\pi)^2} \dfrac{F^{n,n^\prime}_{{\bf k},{\bf q}}(f_{n^\prime,{\bf k}+{\bf q}}-f_{n,{\bf k}})}{\epsilon_{n'{\bf k}+{\bf q}}-\epsilon_{n{\bf k}}-\hbar \omega-i\eta},
\end{equation}
where the prefactor $F^{n,n^\prime}_{{\bf k},{\bf q}}$ denotes the band-overlap of the wavefunctions in Eq.~(\ref{pi2}). Neglecting the local field effect, we can summarize the prefactor of each method as:
\begin{eqnarray}
\nonumber\text{General form}:&&\langle u_{n{\bf k}}|u_{n^\prime,{\bf k}+{\bf q}}\rangle \langle u_{n^\prime,{\bf k}+{\bf q}}|u_{n{\bf k}}\rangle\\
\nonumber\textit{Ab initio}:&&\langle \phi_{n{\bf k}}|e^{-i{\bf q} \cdot {\bf r}}|\phi_{n^\prime,{\bf k}+{\bf q}}\rangle \langle \phi_{n^\prime,{\bf k}+{\bf q}}|e^{i{\bf q} \cdot {\bf r}^\prime}|\phi_{n{\bf k}}\rangle\\
\nonumber\text{Tight-binding}:&&S_{n,{\bf k}}^\dagger S_{n^\prime,{\bf k}+{\bf q}}S_{n^\prime,{\bf k}+{\bf q}}^\dagger S_{n,{\bf k}}\\
\nonumber\text{Dirac}:&&\frac{g_v}{2}(1+nn^\prime \cos\theta)\\
\text{Lindhard}:&&g_v
\end{eqnarray}
where $g_v=2$ is the valley degeneracy for graphene. 
 
\section{Comparison and discussion}

In this section, we compare the polarizability of graphene calculated from different models as discussed above. We treat the parameter-free \textit{ab initio} results as the reference to discuss the validity of each method with different wave-vectors and frequencies. Dependences on chemical potential and temperature are also discussed. It was reported that the temperature dependence of electron transport in graphene is remarkably weak in the temperature range 0.03-300 K \cite{temp1}. So we fix $T =$ 300 K for all calculations except for the discussion on temperature dependence. We assume a rather conservative value for the energy broadening that the damping factor $\eta$ is set to \SI{0.05}{\electronvolt}. This corresponds to the lifetime of electron $\tau\sim10^{-14}$ s for graphene \cite{ilic,abajo1,abajo2}. We numerically solve Eq.~(\ref{pi_tb}), Eq.~(\ref{pi_dirac}), and Eq.~(\ref{pi_lindhard}) to calculate the tight-binding, Dirac, and Lindhard polarizability of graphene, respectively. The nearest neighbor hopping parameter $\gamma$ is set to \SI{2.8}{\electronvolt} that corresponds to a Fermi velocity $v_F = 9.1\times10^5$ m/s. Other analytical expressions from the literature are also calculated and compared.

For the \textit{ab initio} calculation, we start from the ground state calculations by using DFT as implemented in QUANTUM ESPRESSO \cite{qe1,qe2}. Then the \textit{ab initio} polarizability of graphene is calculated on top of the ground state band structure by using the BerkerelyGW package \cite{bgw1,bgw2}. We adapt the norm-conserving pseudopotential generated by the Martins-Troullier approach \cite{martin} with the Perdew-Burke-Ernzerhof (PBE) exchange-correlation functional in the generalized gradient approximation (GGA) \cite{gga}. A plane-wave basis set with 60 Ry energy cut-off is used to expand the Kohn-Sham wave functions. We sample the first Brillouin zone by a $300\times300\times1$ Monkhorst-Pack \cite{mp} grid. The Fermi-Dirac smearing with 0.002 Ry smearing width is adopted to treat the partial occupancies for graphene. The in-plane lattice constant are $a_1 = a_2 =\sqrt{3}a_0=$ \SI{2.46}{\angstrom}. To avoid interactions from the neighboring lattice in the $z$ direction, a large lattice constant of $a_3 = \SI{18}{\angstrom}$ is set to the $z$ direction of the unit cell.

\subsection{Wave-vector}
\begin{figure}
	\includegraphics[width=8.6 cm]{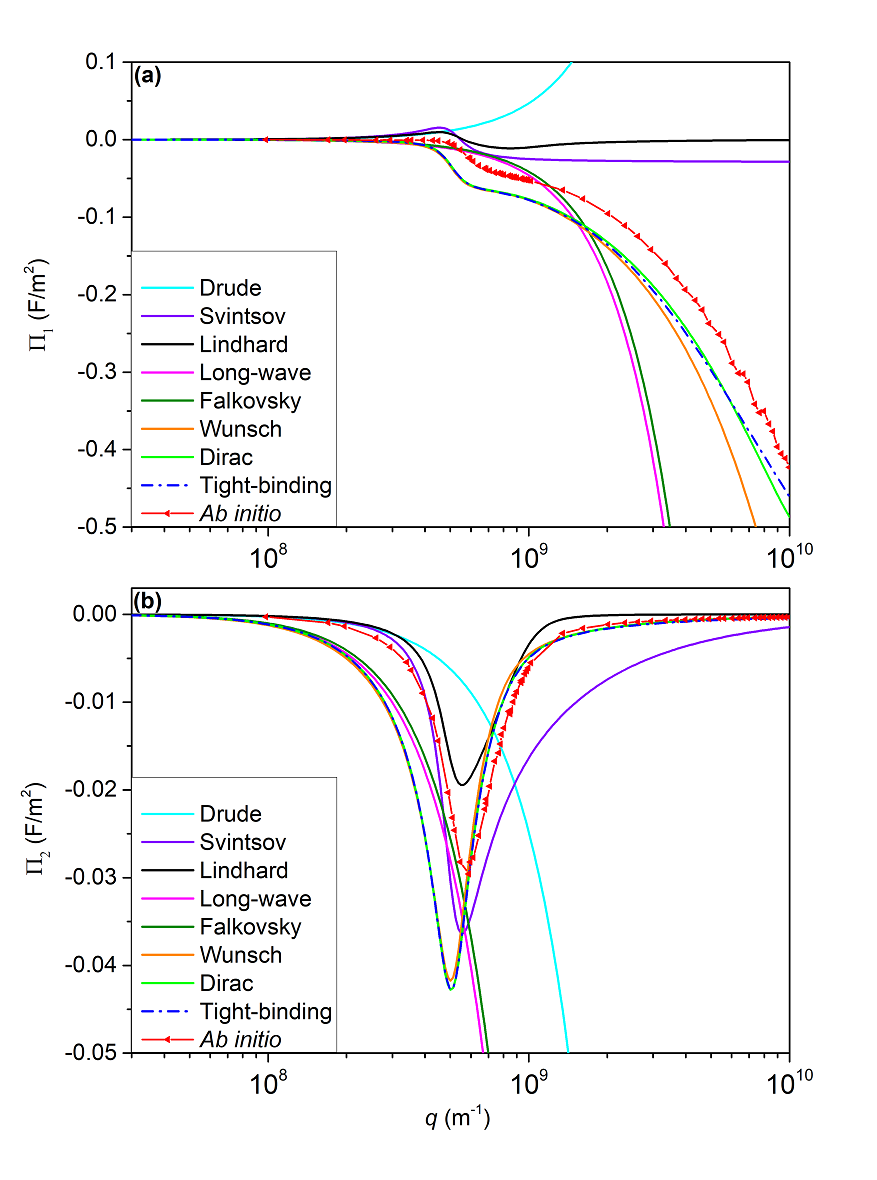}
	\caption{Comparison of real ($\Pi_1$) and imaginary ($\Pi_2$) polarizability of graphene in different wave-vectors. Results of Wunsch and Falkovsky are calculated from the analytical formula given in Ref. \cite{wunsch} and Ref. \cite{falkovsky2}, respectively.}
	\label{fig2}
\end{figure}

We first consider the spatial dispersion of polarizability of graphene. In Fig.~\ref{fig2}, we show the calculated polarizability of graphene obtained from different methods with different wave-vectors. The frequency is fixed at \SI{0.3}{\electronvolt} and the chemical potential $\mu =$ \SI{0.1}{\electronvolt}. Firstly, let us consider the real part polarizability of graphene as shown in Fig.~\ref{fig2}(a). As we can see, when $q<10^8$ $\textrm{m}^{-1}$, the deviation of results from different methods gradually vanish. This agree to the fact that graphene is a Dirac material and all formulas obtained from the Dirac model reduce to the long-wave formula when $v_Fq <<\omega$. Moreover, because the frequency $\hbar\omega$ is not significantly greater than $2\mu$, the intra-band transitions dominate at small wave-vectors so that the Drude polarizability agree well with others. At large wave-vectors, models with only intra-band transitions, i.e., Drude, Svintsov, and Lindhard, failed to describe the spatial non-locality of transport properties of graphene. In particular, the Drude polarizability grows as a positive parabola which presents an unreal screening effect. However, when $q<5\times 10^8$ $\textrm{m}^{-1}$, the Drude polarizability agree well with that of the Lindhard model. This suggests that one can safely use the Drude model to calculate the intra-band contributions in near-field heat transfer when wave-vector is less than 300 times of $\omega/c$ where $c$ is the speed of light. Both Lindhard and Svintsov polarizability saturate when $q>2\times10^9$ $\textrm{m}^{-1}$ that implies the intra-band transitions are negligible for transition with very large wave-vectors. 

Results calculated from Falkovsky's formula and the long-wave formula show a very similar character and they start to deviate from other models when $q > 4\times10^8$ $\textrm{m}^{-1}$. This is because both of them are based on the long-wave approximation that is not supposed to be valid when wave-vectors are large. On the other hand, the tight-binding results agree well with the \textit{ab initio} results except for a global shift. This shift originates from the deviation of tight-binding band structure from the \textit{ab initio} calculations at large wave-vectors. Moreover, the Dirac model produces the same result as the tight-binding model when $q < 5\times10^9$ $\textrm{m}^{-1}$. This implies that the linear dispersion of the Dirac-cone band structure extends up to \SI{3}{\electronvolt}. Besides, polarizability calculated from the Wunsch's formula coincides exactly with those of the tight-binding (and Dirac) model when $q < 2\times10^9$ $\textrm{m}^{-1}$.

Secondly, unlike the real part of polarizability defines the screening of an electromagnetic wave in a medium, the imaginary part is responsible for the absorption of radiation and is proportional to the damping of oscillations. In Fig.~\ref{fig2}(b), similar to the real part, models based on the long-wavelength approximation (Drude, Long-wave, and Falkovsky) are only valid when $q<10^8$ $\textrm{m}^{-1}$ and they are monotonically decreasing with the increase of wave-vectors. The tight-binding, Dirac, and Wunsch model give almost identical values at all $q$ values and they are very close to \textit{ab initio} results. Strong damping peaks are shown at $q\sim 5\times10^8$ $\textrm{m}^{-1}$, which is exactly the resonance of $\hbar v_Fq$ to the frequency $\hbar\omega=$ \SI{0.3}{\electronvolt}. Interestingly, all models with spatial dispersion get a convergent vanishing value when $q>10^9$ $\textrm{m}^{-1}$ that indicates a weak absorption at large wave-vectors.   

\subsection{Frequency}
\begin{figure}
	\includegraphics[width=8.6 cm]{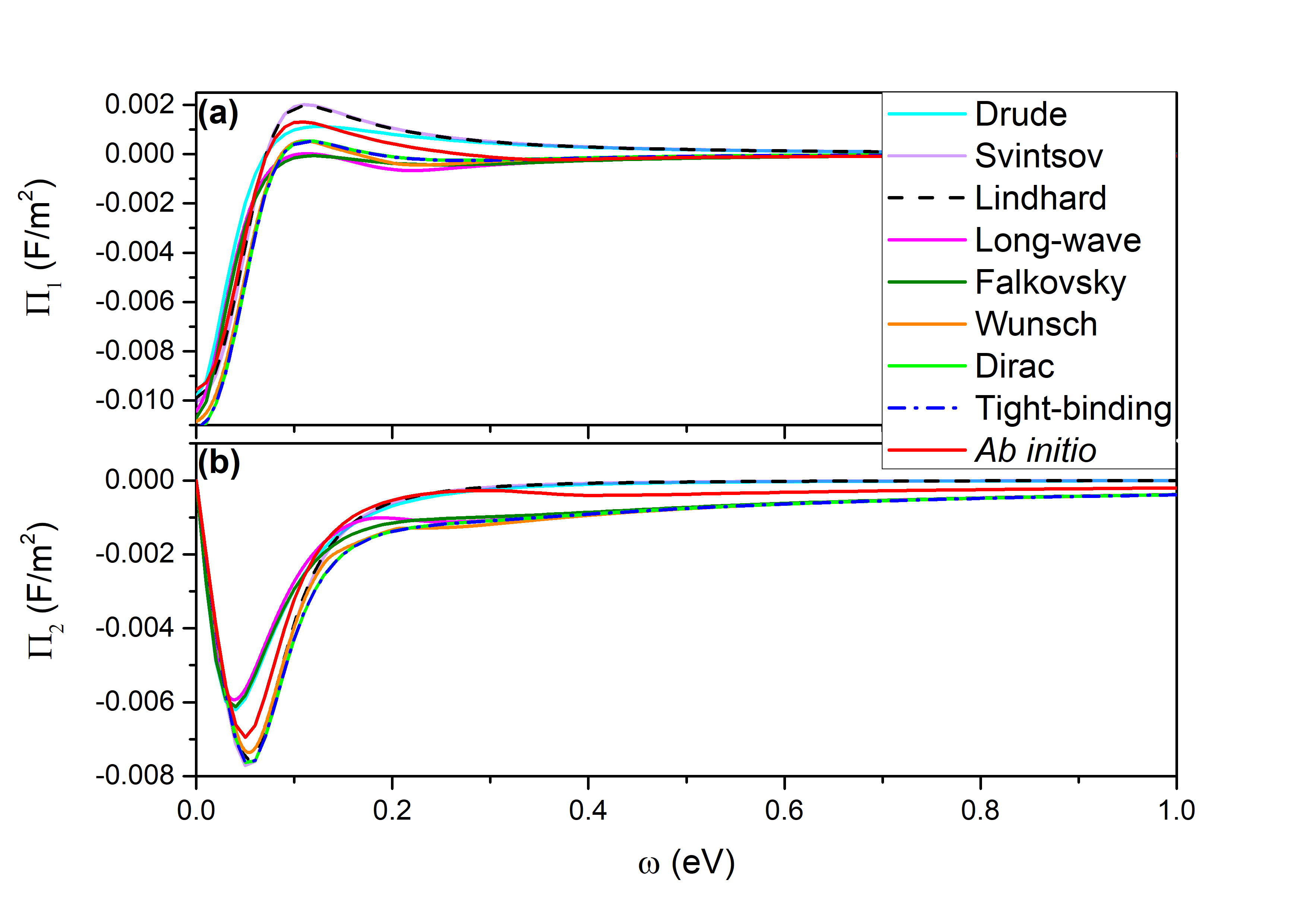}
	\caption{Comparison of real ($\Pi_1$) and imaginary ($\Pi_2$) polarizability of graphene in different frequencies with $q=9.82\times10^7$ $\textrm{m}^{-1}$. Results of Wunsch and Falkovsky are calculated from the analytical formula given in Ref. \cite{wunsch} and Ref. \cite{falkovsky2}, respectively.}
	\label{fig3}
\end{figure}

Next, we discuss the dynamical screening properties of graphene. We first consider the case when the long-wavelength approximation is valid. It is shown in Fig.~\ref{fig2} that when $q<10^8$ $\textrm{m}^{-1}$, the differences between each method are not significant at small chemical potential and low temperature. In Fig.~\ref{fig3} we show the polarizabilities of graphene calculated from different methods with wave-vector fixed at $q=9.82\times10^7$ $\textrm{m}^{-1}$ and chemical potential $\mu =$ \SI{0.1}{\electronvolt}. As shown, all methods produce very similar results. The inter-band transitions make the calculated polarizabilities slightly smaller than those results with only intra-band transitions. The Drude model is surprisingly good when compared with results from \textit{ab initio} calculations, which further supports its validity at low frequencies. All Dirac-type models (Dirac, Long-wave, Falkovsky, and Wunsch) produce almost identical results and consistent with that of the tight-binding model. 

\begin{figure}
 	\includegraphics[width=8.6 cm]{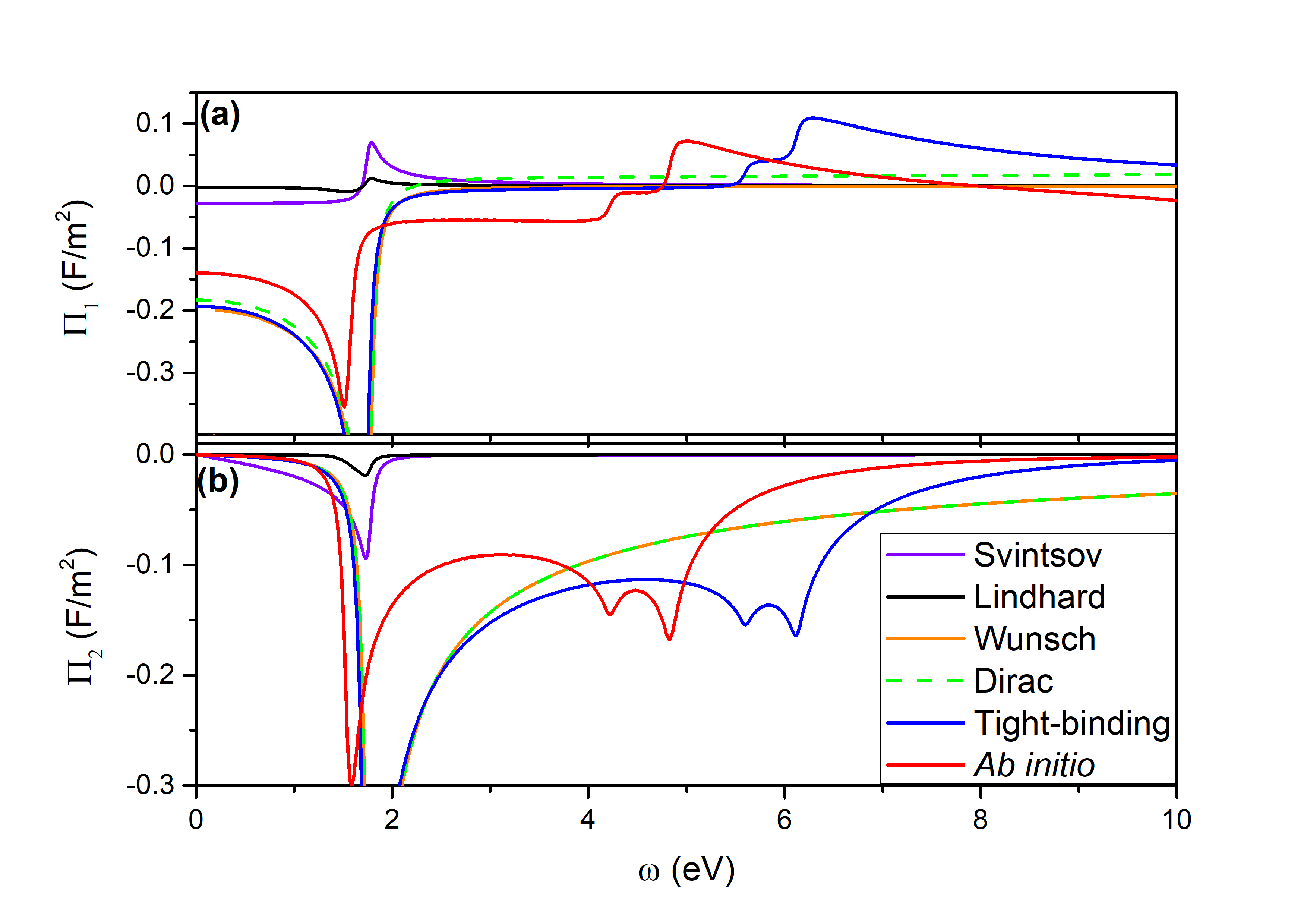}
 	\caption{Comparison of real ($\Pi_1$) and imaginary ($\Pi_2$) polarizability of graphene in different frequencies with $q=2.95\times10^9$ $\textrm{m}^{-1}$. Results of Wunsch is calculated from the analytical formula given in Ref. \cite{wunsch}.}
 	\label{fig4}
 \end{figure}

In Fig.~\ref{fig4}, we compare polarizabilities of graphene obtained from different methods with wave-vector fixed at $q=2.95\times10^9$ $\textrm{m}^{-1}$ and chemical potential $\mu =$ \SI{0.1}{\electronvolt}. With such a large wave-vector, all models based on the long-wave approximation are not valid and we only compare results from models with spatial dispersion. We also extend the frequency window up to \SI{10}{\electronvolt} to show the property of graphene polarizability at high frequencies. As shown in Fig.~\ref{fig4}, the magnitudes of Lindhard and Svintsov polarizability are much smaller than those from other models even in the low-frequency range. This agrees with results as shown in Fig.~\ref{fig2} that intra-band transitions have little contributions at large wave-vectors. 

For models with inter-band transitions, strong peaks are shown around $\sim$ \SI{1.75}{\electronvolt} which resonant to the value of $\hbar v_F q$. When frequencies are smaller than \SI{3}{\electronvolt}, the results of tight-binding, Dirac, as well as Wunsch's model are consistent with each other.
However, when frequencies exceed \SI{3}{\electronvolt}, two prominent absorption peaks are shown in the imaginary part of the \textit{ab initio} polarizability at \SI{4.2}{\electronvolt} and \SI{4.8}{\electronvolt}. These two peaks originate from transitions between the van Hove sigularities around the M points of the Brillouin zone. All models with Dirac dispersion fail to describe this character of polarizability of graphene. This indicates that the linear energy dispersion of $\pi$ band in the Dirac model is not appropriate at high energies. However, except for a global shift, the tight-binding model successfully reproduces the main character of \textit{ab initio} polarizability at high frequencies, agree to previously discussed.

\subsection{Chemical Potential}

\begin{figure}
 	\includegraphics[width=8.6 cm]{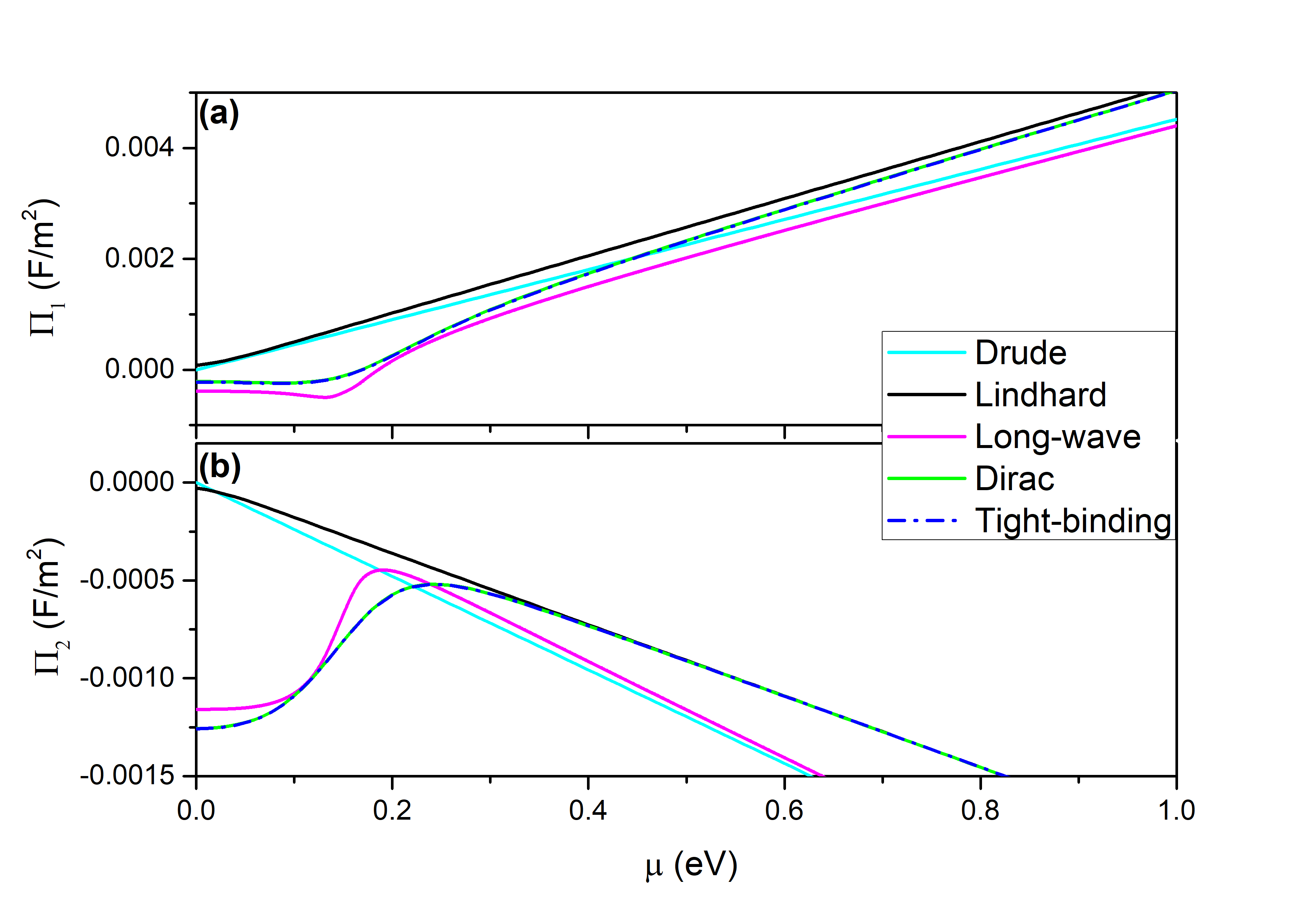}
 	\caption{Comparison of real ($\Pi_1$) and imaginary ($\Pi_2$) polarizability of graphene in different chemical potentials.}
 	\label{fig5}
 \end{figure}

Besides the wave-vector and frequency, effects from doping (chemical potential) and temperature can also play an important role in screening effects of graphene. Figure \ref{fig5} shows the calculated polarizability of graphene with different chemical potentials. The wave-vector is fixed at a small value of $9.82\times10^7$ $\textrm{m}^{-1}$ and the frequency is fixed at \SI{0.3}{\electronvolt}. The chemical potential dependence mainly originates from the Fermi function as shown in Eq.~(\ref{pi2}). However, at zero temperature, the Fermi function becomes a step function as the case in \textit{ab initio} method. In order to tune the chemical potential in \textit{ab initio} method, one need to introduce extra charges in the ground state calculations and it is unstable with large doping. So, in Fig.~\ref{fig5}, we show the chemical potential dependence of graphene polarizability obtained from other models. As we can see, both real and imaginary parts of polarizability have a linear relation to the chemical potential when $\mu \gg \hbar\omega$. Results from the Dirac model coincide exactly with that of the tight-binding model and they are asymptotically approaching results from the Lindhard formula. This implies that the intra-band transitions dominate the screening effect at large chemical potentials. This agrees with the fact that rich electrical doping makes the electronic structure of graphene close to the 2DEG. 

On the other hand, when $\mu<\hbar\omega$, the inter-band transitions contribute and the value of real part polarizability gradually decreases and saturates to a negative value at zero chemical potentials. In this case, the intra-band transitions vanish and the screening effect is from the inter-band transitions so that the system behaves as a dielectric material. 

Similar behaviors are shown for results from models with the long-wave approximation. As shown, the Drude polarizability is also linearly proportional to the chemical potential but with a different slope from the Lindhard results. Because only intra-band transitions are considered, both the Drude and Lindhard model give a vanishing value at zero chemical potential. Again, results from the long-wave formula asymptotically converge to the Drude polarizability when chemical potential is large. 

\subsection{Temperature} 

\begin{figure}
	\includegraphics[width=8.6 cm]{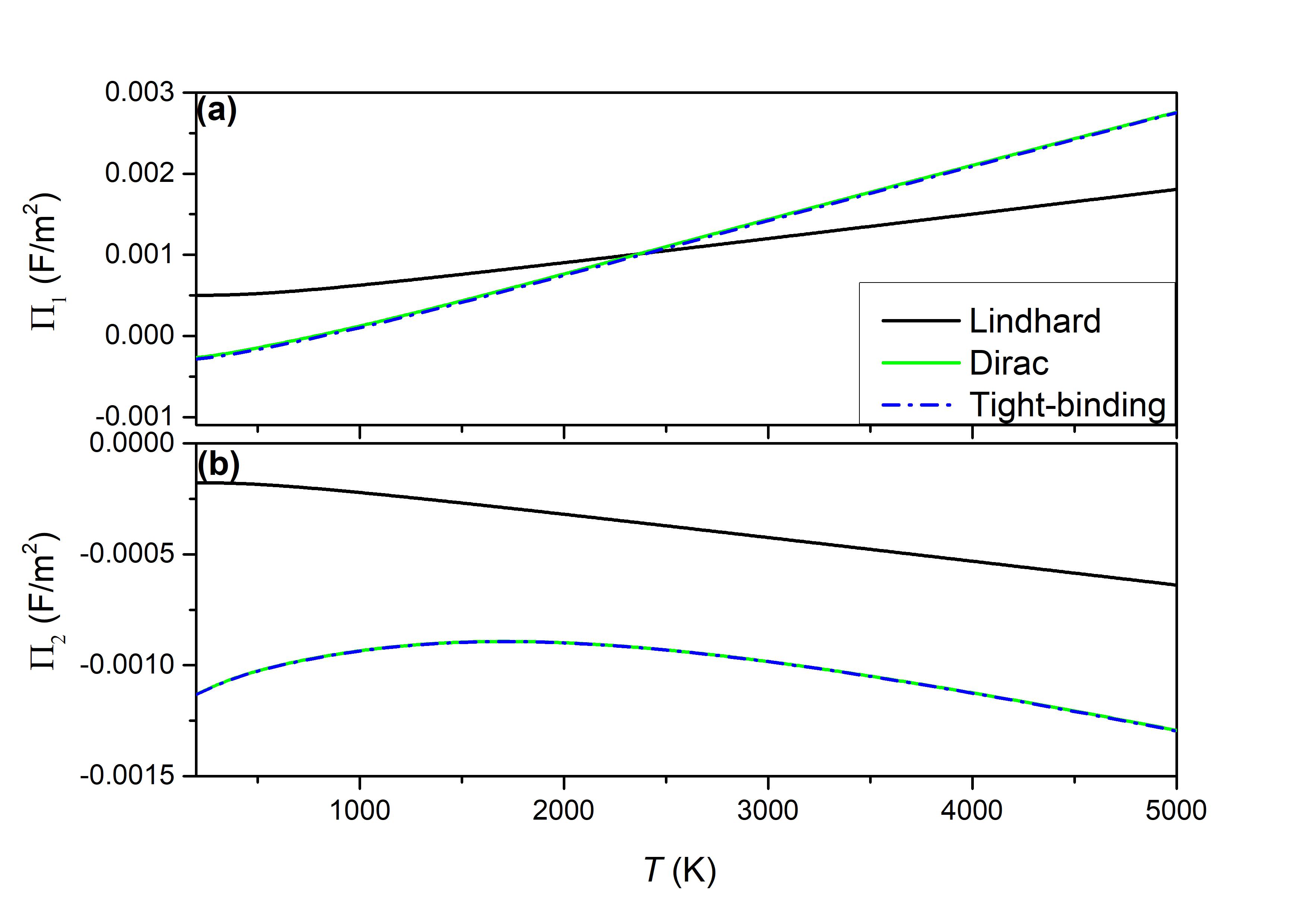}
	\caption{Comparison of real ($\Pi_1$) and imaginary ($\Pi_2$) polarizability of graphene in different temperatures.}
	\label{fig6}
\end{figure}

At last, we discuss the effect of temperature. The \textit{ab initio} method calculates the ground state electronic structure at zero temperature. As reported \cite{temp1}, the temperature dependence is not significant for graphene polarizability when $T<300$ K. In Fig.~\ref{fig6}, we show the calculated polarizability of graphene with temperature ranges from 200 K to 5000 K. The chemical potential was set to \SI{0.1}{\electronvolt} with a fixed wave-vector $q = 9.82\times10^7$ $\textrm{m}^{-1}$ and frequency $\hbar\omega = $ \SI{0.3}{\electronvolt}. As seen, the general feature of temperature dependence is very similar to that of chemical potentials as shown in Fig.~\ref{fig6}. For real part polarizability, both Lindhard, Dirac, and the Tight-binding results show an approximately linear dependence with respect to temperature. The Lindhard polarizability has a smaller slope than that of Dirac and tight-binding results which implies that temperature dependence of intra-band transitions is not as sensitive as that of the inter-band transitions. The imaginary part polarizability, however, is nonmonotonic for results from the Dirac and tight-binding model. It increasing with temperature when $T<1800$ K, and decreases at high temperatures that agree to the previous report \cite{temp2}. The overall temperature dependence is not significant when $T < 3000$ K. 

\section{Conclusion}

In this work, we discussed the dynamical polarizability of graphene with four theoretical approaches. We derive the general expression of polarizability from the linear response Kubo formula. Different methods are distinguished by a prefactor that represents the band-overlap of wavefunctions of unperturbed Hamiltonian. We discussed the validity of each method by comparing their results with different wave-vector, frequency, chemical potential, and temperature. At finite doping, the Lindhard formula for 2DEG describes the intra-band transitions of graphene and it reduces to the Drude formula when both wave-vector and temperature go to zero. For models with inter-band transitions, the tight-binding method produces similar results as the \textit{ab initio} calculation except for a global shift that originates from the deviation of their band structure at high energies. Moreover, the Dirac model is equally good as the tight-binding method when energy is smaller than \SI{3}{\electronvolt}. At zero temperature, all theoretical models reduce to a simple analytical long-wave formula in the long-wavelength limit and its intra-band term corresponds to the Drude conductivity. The intra-band transitions becomes dominant at large electrical doping and all theoretical model give asymptotic results with large chemical potential. The temperature dependence of the real part polarizability of graphene is approximately linear while a non-monotonic temperature dependence is shown for the imaginary part. Our results may provide a solid reference for applications of the response and screening properties of graphene and the methods can be employed to other Dirac materials with appropriate theoretical treatments.

\section{acknowledgments}

We would like to thank Jiebin Peng, Zhibin Gao, and Zu-Quan Zhang for insightful discussions. This work is supported by a FRC grant R-144-000-427-114 and an MOE tier 2 grant R-144-000-411-112.


\vfill

\begin{thebibliography}{[Vo]}
	\bibitem{graphene}
	K. S. Novoselov, A. K. Geim, S. V. Morozov, D. Jiang, Y. Zhang, S. V. Dubonos, I. V. Gregorieva, and A. A. Firsov, Science \textbf{306}, 666 (2004).
    \bibitem{graphene2}
    A. K. Geim, and K. S. Novoselov, Nat. Mater. \textbf{6}, 183-191 (2007).
    \bibitem{graphene3}
    A. K. Geim, Science \textbf{324}, 1530 (2009).
    \bibitem{graphene4}
	A.H. Castro Neto, F. Guinea, N. M. R. Peres, K. S. Novoselov, and A. K. Geim, Rev. Mod. Phys. \textbf{81}, 109 (2009).
    \bibitem{energy1}
    A. I. Volokitin and B. N. J. Persson, Rev. Mod. Phys. \textbf{79}, 1291 (2007).
    \bibitem{energy2} 
    J. -S. Wang and J. Peng, EPL \textbf{118}, 24001 (2017).
    \bibitem{energy3}
    J.-H. Jiang, and J.-S.Wang, Phys. Rev. B \textbf{96}, 155437 (2017). 
    \bibitem{energy4}
    Z.-Q. Zhang, J. T. L\"u, and J.-S. Wang, Phys. Rev. B \textbf{97}, 195450 (2018).
    \bibitem{energy5} 
    R. Yu, A. Manjavacas, and F. J. Garc\'{i}a de Abajo, Nat. Commun. \textbf{8}, 2 (2017).
    \bibitem{energy6}
    T. Zhu, Z. -Q. Zhang, Z. Gao, and J. -S. Wang, Phys. Rev. Appl. \textbf{14}, 024080 (2020).
    \bibitem{energy7}
    B. Zhao, B. Guizal, Z. M. Zhang, S. Fan, and M. Antezza, Phys. Rev. B \textbf{95}, 245437 (2017).
    \bibitem{energy8}
    M. -J. He, H. Qi, Y. -T. Ren, Y. -J. Zhao, and M. Antezza, Appl. Phys. Lett. \textbf{115}, 263101 (2019).
    \bibitem{casimir1}
    G. L. Klimchitskaya, U. Mohideen, V. M. Mostepanenko, Rev. Mod. Phys. \textbf{81}, 1827 (2009). 
    \bibitem{casimir2}
    D. Drosdoff and Lilia M. Woods, Phys. Rev. B \textbf{82}, 155459 (2010).
    \bibitem{casimir3}
    V. Yannopapas and N. V. Vitanov, Phys. Rev. Lett. \textbf{103}, 120401 (2009).
    \bibitem{casimir4}
    W. -K. Tse and A. H. MacDonald, Phys. Rev. Lett. \textbf{109}, 236806 (2012).
    \bibitem{casimir5}
    C. Abbas, B. Guizal, and M. Antezza, Phys. Rev. Lett. \textbf{118}, 126101 (2017).
    \bibitem{casimir6}
    B. Guizal and M. Antezza, Phys. Rev. B \textbf{93}, 115427 (2016).
    \bibitem{wunsch}
    B. Wunsch, T. Stauber, F. Sols, and F. Guinea, New J. Phys. \textbf{8}, 318 (2006).
    \bibitem{huang}
    E. H. Hwang and S. Das Sarma, Phys. Rev. B \textbf{75}, 205418 (2007).
    \bibitem{gusynin}
    V. P. Gusynin, S. G. Sharapov, and J. P. Carbotte, Phys. Rev. Lett. \textbf{96}, 256802 (2006).
    \bibitem{gusynin2}
    V. P. Gusynin, S. G. Sharapov and J. P. Carbotte, Int. J. Mod. Phys. B \textbf{21}, 4611 (2007).  
    \bibitem{falkovsky}
    L. A. Falkovsky and A. A. Varlamov, Eur. Phys. J. B \textbf{56}, 281-284 (2007).
    \bibitem{falkovsky2}
    L. A. Falkovsky, J. Phys.: Conf. Ser. \textbf{129}, 012004 (2008).
    \bibitem{klimchitskaya}
    G. L. Klimchitskaya and V. M. Mostepanenko, Phys. Rev. B \textbf{94}, 195405 (2016).
    \bibitem{klimchitskaya2}
    G. L. Klimchitskaya, V. M. Mostepanenko, and V. M. Petrov, Phys. Rev. B \textbf{96}, 235432 (2017).
    \bibitem{bordag1}
    M. Bordag, G. L. Klimchitskaya, V. M. Mostepanenko, and V. M. Petrov, Phys. Rev. D \textbf{91}, 045037 (2015).
    \bibitem{bordag2}
    M. Bordag, I. Fialkovskiy, and D. Vassilevich, Phys. Rev. B \textbf{93}, 075414 (2016).
    \bibitem{mahan}
    G. D. Mahan, \textit{Many-Particle Physics}, (Springer, New York, 2000).
    \bibitem{lindhard}
    J. Lindhard, K. Dan. Vidensk. Selsk. Mat. Fys. Medd. \textbf{28}, 8 (1954). 
    \bibitem{kubo}
    R. Kubo, J. Phys. Soc. Jpn. \textbf{12}, 570 (1957),\\
    R. Kubo, M. Yokota, and S. Nakajima, J. Phys. Soc. Jpn. \textbf{12}, 1203 (1957).
    \bibitem{dyson}
    F. J. Dyson, Phys. Rev. \textbf{75}, 1736 (1949).
    \bibitem{Onida}
    G. Onida, L. Reining, and A. Rubio, Rev. Mod. Phys. \textbf{74}, 601-659 (2002).
    \bibitem{rpa}
    H. Ehrenreich, M. H. Cohen, Phys. Rev. \textbf{115}, 786 (1959).
    \bibitem{gw}
    F. Aryasetiawan, and O. Gunnarsson, Rep. Prog. Phys. \textbf{61}, 237 (1998).
    \bibitem{ks}
    W. Kohn, and L. Sham. Phys. Rev. \textbf{140}, A1133 (1965).
    \bibitem{slater}
    J. C. Slater, Phys. Rev. \textbf{34}, 1293 (1929).
    \bibitem{adler}
    S. L. Adler, Phys. Rev. \textbf{126}, 413 (1962).
    \bibitem{wiser}
    N. Wiser, Phys. Rev. \textbf{129}, 62 (1963).
    \bibitem{bse}
    E. E. Salpeter and H. A. Bethe, Phys. Rev. \textbf{84}, 1232 (1951).
    \bibitem{lfe}
    T. Zhu, P. E. Trevisanutto, T. C. Asmara, L. Xu, Y. P. Feng, and A. Rusydi, Phys. Rev. B \textbf{98}, 235115 (2018). 
    \bibitem{lfe1}
    S. G. Louie, J. R. Chelikowsky, and M. L. Cohen, Phys. Rev. Lett. \textbf{34}, 155 (1975).
    \bibitem{mermin}
    N. D. Mermin, Phys. Rev. B \textbf{1}, 2362 (1970).
    \bibitem{dassarma}
    S. Das Sarma and E. H. Hwang, Phys. Rev. B \textbf{54}, 1936 (1996).
    \bibitem{drude}
    J. Horng, C. -F. Chen, B. Geng, C. Girit, Y. Zhang, Z. Hao, H. A. Bechtel, M. Martin, A. Zettl, M. F. Crommie, Y. R. Shen, and F. Wang, Phys. Rev. B \textbf{83}, 165113 (2011).
    \bibitem{Lewkowicz}
    M. Lewkowicz and B. Rosenstein, Phys. Rev. Lett. \textbf{102}, 106802 (2009). 
    \bibitem{svintsov}
    D. Svintsov and V. Ryzhii, Phys. Rev. Lett. \textbf{123}, 219401, (2019).
    \bibitem{temp1}
    Y.-W. Tan, Y. Zhang, H. L. Stormer, and P. Kim, Eur. Phys. J. Special Topics \textbf{148}, 15–18 (2007).
    \bibitem{ilic}
    O. Ilic, M. Jablan, J. D. Joannopoulos, I. Celanovic, H. Buljan, and M. Solja\v{c}i\'{c}, Phys. Rev. B \textbf{85}, 155422, (2012).
    \bibitem{abajo1}
    F. J. Garc\'{i}a de Abajo, ACS Nano, \textbf{7} 11409 (2013).
    \bibitem{abajo2}
    J. D. Cox and F. J. Garc\'{i}a de Abajo, Nat. Commun. 5 5725 (2014).  
    \bibitem{qe1}
    P. Giannozzi \textit{et al.}, J.Phys.:Condens.Matter \textbf{21}, 395502 (2009).
    \bibitem{qe2}
    P. Giannozzi \textit{et al.}, J.Phys.:Condens.Matter \textbf{29}, 465901 (2017).
   \bibitem{bgw1}
    M. S. Hybertsen and S. G. Louie, Phys. Rev. B \textbf{34}, 5390 (1986).
   \bibitem{bgw2}
   J. Deslippe, G. Samsonidze, D. A. Strubbe, M. Jain, M. L. Cohen, and S. G. Louie, Comput. Phys. Commun. \textbf{183}, 1269 (2012).  
   \bibitem{martin}
   N. Troullier and J. L. Martins, Phys. Rev. B \textbf{43}, 1993 (1991).
   \bibitem{gga}
   J. P. Perdew, K. Burke, and M. Ernzerhof, Phys. Rev. Lett. \textbf{77}, 3865 (1996).
   \bibitem{mp}
   H. J. Monkhorst, and J. D. Pack, Phys. Rev. B \textbf{13}, 5188 (1976).
   \bibitem{temp2}
   E. H. Hwang and S. Das Sarma, Phys. Rev. B \textbf{79}, 165404 (2009).
\end{thebibliography}
\end{document}